\documentclass[twocolumn,showpacs,amsmath,amssymb,prl]{revtex4}
\usepackage{bm}% bold math
\begin{document}

\title{Vortex avalanches and the onset of superfluid turbulence}
\author{N. B. Kopnin}

\affiliation{Low Temperature Laboratory, Helsinki University of
Technology, P.O. Box 2200, FIN-02015 HUT, Finland,\\
L. D. Landau Institute for Theoretical Physics, 117940 Moscow,
Russia}

\date{\today}

\begin{abstract}
Quantized circulation, absence of Galilean invariance due to a
clamped normal component, and the vortex mutual friction are the
major factors that make superfluid turbulence behave in a way
different from that in classical fluids. The model is developed
for the onset of superfluid turbulence that describes the initial
avalanche-like multiplication of vortices into a turbulent vortex
tangle.
\end{abstract}

\pacs{67.57.Fg, 67.40.Vs, 47.37.+q, 47.37.+q, 47.32.Cc}

\maketitle

Turbulence in classical liquids \cite{Landau/Hydro} is governed by
the interplay of the inertial (the first) and viscous (the second)
terms in the r.h.s. of the Navier-Stokes equation:
\begin{equation}
\frac{\partial {\bf v}}{\partial t}+ \bm{\nabla }\tilde \mu = {\bf
v}\times
 \bm{\omega}+  \nu\nabla^2 {\bf v} \ , \label{NormalHydrodynamics}
\end{equation}
where $\bm{\omega}=\bm{\nabla}\times {\bf v}$, $\tilde \mu =\mu
+v^2/2$, and $\mu$ is the chemical potential in the case of an
isothermal flow. The transition to turbulence is determined by the
Reynolds number ${\rm Re}=RU/\nu$ formed by the characteristic
velocity $U$ of the flow and the characteristic size $R$ of the
system together with the kinematic viscosity $\nu$. For small
Reynolds numbers, the viscous term $-\nu k^2 {\bf v}_{\bf k}$ for
a perturbation with a wave vector ${\bf k}$ stabilizes the laminar
flow. On the contrary, at large ${\rm Re}\gg 1$ the effect of the
inertial term in Eq.\ (\ref{NormalHydrodynamics}) is dominating,
and laminar flow becomes increasingly unstable towards formation
of a chaotic flow of eddies. The evolution of turbulence is
described by the Kolmogorov energy cascade: the kinetic energy of
the flow is transferred to smaller and smaller length scales, via
decay into smaller vortex loops along the Richardson cascade
\cite{Frisch}, until a scale is reached where the energy is
dissipated by viscosity.

In superfluids turbulence acquires new features: Firstly, because
it consists of two inter-penetrating components, i.e., the
frictionless superfluid and the viscous normal fractions, and
secondly, because the vorticity of the superfluid component is
quantized in terms of the elementary circulation quantum $\kappa
=2\pi \hbar /M$ where $M$ is a mass of a ``superfluid particle''
(for $^3$He, it is the mass of a pair of atoms, $M=2m$). If both
the normal and the superfluid components are moving, the turbulent
state bears more resemblance to the turbulence of classical
viscous liquids. The superfluid turbulence is now a subject of
intensive studies, see \cite{VinenNiemela} for a review. A new
class of turbulent flow becomes possible when the normal component
is so viscous that it is essentially immobile while the main flow
is supported by the superfluid component which contains a large
number of quantized vortices. We refer to this state as {\it
one-component superfluid turbulence}. A superfluid motion of such
type can be realized for flows in narrow channels and in rotating
superfluids. The velocity of superflow can be characterized by the
``superfluid Reynolds number'' ${\rm Re}_s=U_sR/\kappa$ where
$U_s$ is the mean superfluid velocity with respect to the normal
component (counterflow velocity). The condition ${\rm Re}_s\sim 1$
corresponds to the Feynman criterion at which it becomes
energetically favorable to unpin a vortex from the container wall.
If a large nucleation barrier for vortex nucleation exists, then
vortices are not necessarily created even at high superfluid
Reynolds numbers ${\rm Re}_s>1$ and the superfluid remains in a
vortex-free counterflow state. For velocities well below the
intrinsic critical velocity\cite{crit-vel} $v_c\sim \kappa /\xi$,
superfluid turbulence can be initiated if quantized vortices are
injected by some extrinsic means into the flow. The turbulence
develops when initial vortices start to multiply and form a vortex
tangle. Recent experiments in rotating $^3$He-B \cite{Nature03}
have revealed qualitatively new conditions for the onset of
one-component superfluid turbulence: In contrast to the normal
fluids, the superfluid turbulence develops below certain
temperature independent of the rotation velocity provided the
latter is sufficiently high. According to the measurements of
Ref.\ \cite{Nature03}, a few seed vortices injected into the
$^3$He B counterflow triggered a transition into a state with a
turbulent vortex tangle for temperatures $T<0.6 T_c$ while the
same injection at $T>0.6T_c$ did not create any substantial number
of vortices in the final state. In these measurements, Re$_s$ as
high as 200 have been reached without any noticeable dependence of
the transition temperature on the initial counterflow velocity.
The theoretical arguments supported by numerical simulations
explain \cite{Nature03} these conditions in terms of the mutual
friction between the normal and superfluid components that appears
in the presence of quantized vortices.

In this Letter we develop a model that describes the onset of
one-component superfluid turbulence employing the ideas put
forward in Ref.\ \cite{Nature03}. We consider here the initial
multiplication of vortex loops immediately after injection of seed
loops into a superflow with Re$_s \gg 1$. Since superfluid
vorticity is quantized, formation of vortices during the onset of
turbulence is the key issue. We consider this process taking into
account effects of mutual friction and derive an equation for
evolution of the vortex loop density at the initial stage of the
transition to turbulence.

Quantized vortices mediate transfer of momentum between the normal
and superfluid components which produces the mutual friction force
on a unit volume of superfluid \cite{HallVinen,Sonin}
\begin{equation}
{\bf f}_{\rm mf} =-\alpha ^\prime \rho _{\rm s} [{\bf v}_{\rm
s}\times \bm{\omega}_{\rm s}]+\alpha \rho_{\rm
s}[\hat{\bm{\omega}}_{\rm s}\times [\bm{\omega}_{\rm s} \times
{\bf v}_s]] \label{mutfricforce}
\end{equation}
for a normal component stationary in the container frame, ${\bf
v}_n=0$. Here $\bm{\omega}_s=\bm{\nabla} \times {\bf v}_s$ and
$\hat {\bm{\omega}}_s$ is the unit vector. For an array of
quantized vortices
\[
\bm{\omega}_s ={\kappa} \sum_n  \int d{\bf r}_n\, \delta
\left({\bf r}-{\bf r}_n\right)
\]
where ${\bf r}_n$ is the coordinate of the $n$-th vortex line, and
$\delta \left({\bf r}-{\bf r}_n\right)$ is a three-dimensional
$\delta$-function. The dimensionless parameters $\alpha$ and
$\alpha'$ describe the mutual friction between the superfluid and
normal components. The mutual friction typically arises from
interaction between quasi-particle excitations and the vortices
\cite{Sonin,Kopnin/review}. The energy dissipation is determined
by the viscous mutual friction coefficient $\alpha$ in Eq.\
(\ref{mutfricforce}), and $\alpha'$ is the reactive coefficient.
Including the mutual friction force in the Euler equation for
superfluid velocity we arrive at \cite{Sonin}
\begin{equation}
\frac{\partial {\bf v}_{\rm s}}{ \partial t}+ \bm{\nabla}\tilde
\mu = (1-\alpha'){\bf v}_{\rm s} \times \bm{\omega}_s +
\alpha~\hat{\bm{\omega}}_s \times\left(\bm{\omega}_s \times{\bf
v}_{\rm s}\right)  \label{SuperfluidHydrodynamics}
\end{equation}
where $\alpha,\, 1-\alpha ^\prime >0$.

Regarding the onset of turbulence, the reactive coefficient
$\alpha' $ in Eq.\ (\ref{SuperfluidHydrodynamics}) simply
renormalizes the inertial term of conventional hydrodynamics [the
first term in r.h.s. of Eq.\ (\ref{NormalHydrodynamics})]. Indeed,
performing the stability analysis of Eq.\
(\ref{SuperfluidHydrodynamics}), we would write ${\bf v}_s ={\bf
v}_{s0}+{\bf v}_s^\prime$, etc., where ${\bf v}_{s0}$ is the
solution of a stationary problem while ${\bf v}_{s}^\prime (t)$ is
a time-dependent perturbation and obtain a first-order in time
linear differential equation  for ${\bf v}_{s}^\prime (t)$ with
time-independent coefficients made out of ${\bf v}_{s0}$ and its
spatial derivatives. Making re-scaling of time $t=t^\prime
/(1-\alpha ^\prime)$ we arrive at the linearized version of Eq.\
(\ref{SuperfluidHydrodynamics}) where the first term has now the
prefactor unity like the one in the Euler equation [i.e., in Eq.\
(\ref{NormalHydrodynamics}) without the viscous term]. The
coefficient $\alpha $ in front of the dissipative term [the second
term in the r.h.s. of Eq.\ (\ref{SuperfluidHydrodynamics})] is
replaced with $q=\alpha/(1-\alpha')>0$. Therefore, one concludes
that the first term in the r.h.s. of Eq.\
(\ref{SuperfluidHydrodynamics}) drives the flow instability
towards turbulence in the same way as the inertial term in the
Euler equation does for a potential flow in classical
hydrodynamics \cite{Landau/Hydro}. The fundamental difference from
classical turbulence is that the dissipative term in Eq.\
(\ref{SuperfluidHydrodynamics}) that stabilizes the flow has now
the same scaling dependence on velocity and its gradients as the
inertial term. The relative importance of dissipation in formation
of turbulence is thus described by the intrinsic dimensionless
parameter of the superfluid, $q=\alpha/(1-\alpha')$ independent of
extrinsic quantities such as $U_s$ or $R$.

The instability leading to multiplication of seed vortices can be
derived from Eq.\ (\ref{SuperfluidHydrodynamics}). Taking curl of
the both sides of Eq.\ (\ref{SuperfluidHydrodynamics}) we obtain
the vorticity equation
\begin{equation}
\frac{\partial \bm{\omega}_s}{\partial t}=(1-\alpha ^\prime
)\bm{\nabla}\times [{\bf v}_s\times \bm{\omega}_s]+\alpha
\bm{\nabla}\times [\hat{\bm{\omega}}_s\times [\bm{\omega}_s\times
{\bf v}_s]] \ . \label{vorticity}
\end{equation}
Let $\ell$ be a characteristic size of vortex loops. In an
entangled vortex-loop state, their three-dimensional density is $
n \sim \ell ^{-3} $ while the vortex-loop length per unit
volume\cite{Vinen1957} (two-dimensional vortex density) is $
L=\ell n=\ell ^{-2}=n^{2/3} $. Let us express the two terms in the
r.h.s. of Eq.\ (\ref{vorticity}) through the vortex density $L$.
Since $\omega _s\sim \kappa L$, the first term in the r.h.s. that
drives the instability becomes
\begin{equation}
\dot L_+ \sim (1-\alpha ^\prime )v_s L^{3/2} \sim (1-\alpha
^\prime )(U_{s}-v_0) L^{3/2} \ . \label{creation}
\end{equation}

The superfluid velocity in Eq.\ (\ref{creation}) is assumed to be
${v}_s={ U}_{s}-{v}_0$ where $U_{s}$ is the counterflow velocity,
and $ v_0\sim \kappa /\ell$ is the self-induced velocity for a
vortex loop of a length $\ell$. The kinetic energy of superfluid
grows due to the increase in the loop density. The kinetic energy
is taken from the external source at the length scale $R$ and is
transferred to smaller scales. The energy flow into the unit
volume of the fluid is $dE/dt \sim E_L \dot L_+$ where $E_L$ is
the energy of the vortex per unit length. However, the energy
transfer goes on only during the transient period when the
vortex-loop density increases. When the loop density reaches a
value such that $U_{s} =v_0\sim \kappa L^{1/2}$ the vortex
multiplication saturates and the energy transfer stops. If the
density happens to become larger, it will decrease towards
$L_{sat}\sim (U_{s}/\kappa )^2$ while the kinetic energy is
returned from smaller to larger scales and, finally, back to the
external source. In other words, saturation is reached when the
``turbulent superfluid Reynolds number'' ${\rm Re}_{s}^{({\rm
turb})}=U_{s}\ell /\kappa $ is of the order of unity. The
condition Re$_s \gg 1$ ensures the separation of scales $\ell \ll
R$ that is required for formation of a vortex tangle. The limit of
large Re$_s \propto \kappa ^{-1}$ is equivalent to vanishing
Planck constant since $\kappa \propto \hbar $ so that the
vorticity becomes a continuous variable like in classical fluids.
However, the saturation takes place only in quantum superfluids:
$L_{sat}$ diverges for classical fluids where $\kappa $ can
vanish. This difference can be seen in the behavior of the
turbulent velocity in normal fluids \cite{Landau/Hydro}: the
velocity at a length scale $\lambda$ decreases as $v_\lambda
\propto \lambda ^{1/3}$ according to the Kolmogorov scaling law.
On the contrary, the velocity in a superfluid turbulence scales as
$v_{\ell}\sim \kappa /\ell$ for smaller $\ell$. This shows that
the ``effective circulation'' in normal fluid decreases, $\kappa
_{eff} \propto \lambda ^{4/3}$. The lower limit on $\lambda$ in
classical fluids is determined by viscosity at the Kolmogorov
dissipation scale $\lambda _0\sim R/{\rm Re}^{1/4}$
\cite{Landau/Hydro}.

The multiplication of vortex loops described by Eq.\
(\ref{creation}) can be understood in terms of vortex collisions
and interconnections. Such processes were indeed seen in numerical
simulations on quantized vortices \cite{Nature03,Schwarz,Tsubota}.
Reconnection of vortices accompanied by formation of vortex tangle
in normal fluids were considered recently in Refs.\
\cite{Kivotides1,Kivotides2}. Each reconnection of quantized
vortices takes a microscopic time of the order of the
quasiparticle collision time, much shorter than the times involved
in hydrodynamic processes. It is accompanied by a small
dissipation within a volume $\xi ^3$ where $\xi$ is the core
radius of the order of the superfluid coherence length. We
consider these processes as instantaneous and neglect the
corresponding dissipation. The rate of increase in the vortex loop
density should be quadratic in $n$ thus $ \dot n_+ = A v_r n^2
\ell ^2 $. Here $v_r$ is the relative velocity of the vortex
loops, $\ell ^2$ is the loop cross section, a constant $A\sim 1$
describes an ``efficiency'' of the vortex multiplication due to
pair collisions. Using definition of $L$ the vortex multiplication
rate becomes $\dot L_+ \sim v_r L^{3/2}$. The vortex velocity is
determined through the mutual friction parameters $\alpha$ and
$\alpha ^\prime$ such that $ {\bf v}_L=(1 -\alpha ^\prime ){\bf
v}_s -\alpha \, \hat{\bm{\omega}}_s\times {\bf v}_s $. The
relative velocity of loops $v_r$ is proportional to the
longitudinal component of ${\bf v}_L$: $v_r\sim (1-\alpha ^\prime
)v_s$ which agrees with Eq.\ (\ref{creation}).

We assume that vortex multiplication occurs first within a certain
region in the fluid near the location of the seed loops from where
the created vortex tangle penetrates into the rest volume of the
fluid. The effect of the second (viscous) term in the r.h.s. of
Eq.\ (\ref{vorticity}) is to decrease the loop density in the
multiplication region by inflating the loops due to counterflow
and extracting them from this region. The viscous component of the
mutual friction force leads to variation in the vortex loop length
$ \dot \ell \sim 2\pi v_L \sim \alpha (U_{s}-v_0) $. The length
increases while the density decreases if $U_{s} -v_0>0$. When
$v_0>U_{s}$, the density increases since the loops shrink due to
the friction. Thus the rate of vortex-loop density variation due
to the viscous component becomes
\begin{equation}
\dot L_- \sim -\alpha (U_{s}-v_0) L ^{3/2} \ .  \label{inflation}
\end{equation}
The loops expanding according to Eq.\ (\ref{inflation}) are
extracted from the region of their multiplication into the bulk if
$U_s>v_0$ or vice versa if $U_s<v_0$. This process is accompanied
by dissipation of the kinetic energy. The dissipation persists
until the equilibrium density is reached.

As we see, both inertial and viscous mutual friction terms, Eqs.\
(\ref{creation}) and (\ref{inflation}), have the same dependence
on the vortex density, i.e., on the vortex length scale. The total
variation of the loop density in the multiplication region is the
sum of the two processes, $\dot L =\dot L_+ +\dot L_-$. Putting
$v_0=\kappa /\ell =\kappa L^{1/2} $ we obtain
\begin{equation}
\dot L=\beta \left[ U_{s} L^{3/2}-\kappa L^2\right]
\label{eq-Vinen1}
\end{equation}
where $\beta = A(1-\alpha ^\prime )-\alpha =(A-q)/(1-\alpha
^\prime)$.

Equation (\ref{eq-Vinen1}) looks like the Vinen equation
\cite{Vinen1957} for superfluid turbulence. However, the
difference is that the coefficient $\beta$ can now have either
positive or negative sign depending on the mutual friction
parameters. One distinguishes two limits. In the viscosity
dominated regime when $\beta <0$, the rate of extraction of vortex
loops exceeds the rate of multiplication; there is no time for
vortices to multiply since all the seed and newly created vortices
are immediately wiped away into the bulk fluid. The number of
vortices in the final state is essentially equal to the number of
initial vortices, and the turbulent state is not formed. The
corresponding stable solution to Eq.\ (\ref{eq-Vinen1}) is
$L\rightarrow 0$. Equation (\ref{eq-Vinen1}) has another point of
attraction, $L\rightarrow \infty$. This would correspond to a
decay of initially created vortex tangle in a situation without
net counterflow: the vortex loops shrink due to the viscous mutual
friction force.

In the low-viscosity regime when $\beta >0$, the rate of
multiplication is faster, and the number of created vortex loops
is large: Each new vortex loop serves as a source for producing
more vortices. As a result, an avalanche-like multiplication takes
place, which leads to formation of a turbulent vortex tangle. As
the number of vortex loop grows, the self-induced velocity
increases and finally the saturated density $L_{sat}$ is reached.
Indeed, Eq.\ (\ref{eq-Vinen1}) has the stable solution
$L_{sat}=(U_{s}/\kappa )^2$ which corresponds to the Vinen
equilibrium vortex density for the counterflow turbulence
\cite{Schwarz1}. For $\beta >0$, the solution to an equation of
the type of Eq.\ (\ref{eq-Vinen1}) for a constant $U_{s}$ was
found in Ref.\ \cite{Schwarz1}. At the initial stage, when $L\ll
L_{sat}$, the vortex density in the multiplication region varies
as
\begin{equation}
L/L_{sat}=\left[ \left(L_{sat}/L_0\right)^{1/2}-(t/\tau
)\right]^{-2}  \label{evolution}
\end{equation}
with the rate
\[
\tau ^{-1}=\beta U_{s}^2/2\kappa \ .
\]
In Eq.\ (\ref{evolution}), $L_0$ is the (small) initial vortex
density at $t=0$. The time dependence of $L$ turns to exponential
$L_{sat}-L \propto \exp (-t/\tau )$ for $L\rightarrow L_{sat}$.
The time $\tau $ decreases with lowering the temperature since
$\beta$ increases. The energy $E_L\dot L_+$ pumped into the fluid
goes first to a reversible increase in the kinetic energy of the
superflow. A part of it is then dissipated due to the mutual
friction.

The overall evolution of the vortex density can be seen as an
interplay of two processes. The first is the turbulent instability
governed by Eq.\ (\ref{creation}). The second process is expansion
of vortex loops from the multiplication region into the bulk due
to the viscous part of the mutual friction force. Equation
(\ref{inflation}) taken with the {\it opposite} sign gives the
rate of vortex-loop-density flow {\it  into the bulk}. In this
form, it coincides exactly with the Vinen equation
\cite{Vinen1957} as derived by Schwarz \cite{Schwarz}. The
solution for a constant $U_{s}$ has the form of Eq.\
(\ref{evolution}) with the characteristic rate
\[
\tau _b^{-1}= \alpha U_{s}^2/2\kappa \ .
\]
This time increases with lowering the temperature since $\alpha$
decreases.

The instability develops out of initial seed vortices introduced
into the fluid. In experiments \cite{Nature03} these seed vortices
were produced by the Kelvin--Helmholtz instability at the
interface between the A and B phases in rotation \cite{KHinstab}.
We note that the inequality $q\ll 1$ is also the condition for
propagation of the Kelvin waves \cite{Barenghi,Vinen2003}.
Perturbations of the seed vortex lines in the form of Kelvin waves
can grow and multiply under the action of longitudinal flows.
These processes obviously belong to the same class and can also be
described within our model.

Consider a container of a radius $R$ rotating with an angular
velocity $\Omega$. If the counterflow velocity were kept constant
$U_{s}=\Omega R$ as in the vortex-free state, the density
$L_{sat}$ would correspond to the number of vortices in the
container $N_{max}\sim \pi R^2L_{sat}\sim N_{eq}^2$ where
$N_{eq}=2\pi R^2\Omega /\kappa$ is the equilibrium number of
vortices. Such a big number $N_{max}$ of vortices is unstable,
therefore it relaxes towards $N_{eq}$. This decay of turbulence
takes much longer times than $\tau$ and is accompanied by
polarization of vortex loops and screening of the counterflow,
which is beyond the scope of the present paper. In the final
state, with the vortex cluster formed in the center of the
container, the counterflow velocity is decreased down to
$U_{s}\sim \kappa /r_0\sim (\kappa \Omega )^{1/2}$ where $r_0$ is
the intervortex distance for the equilibrium density $L_{eq}\sim
\Omega /\kappa$.

In Fermi superfluids and superconductors, the parameter $q\approx
(\omega_0\tau)^{-1}$, where $\omega_0$ is the spacing between the
bound states of quasiparticles in the vortex core and $\tau^{-1}$
is their lifetime  broadening due to scattering from the normal
component \cite{Kopnin/review}. The ratio $\alpha /(1-\alpha
^\prime)$ decreases exponentially
\cite{KL2tau,Kopnin/review,Bevan} at low temperatures. As a
result, below a certain temperature, an avalanche-like burst of
vortices in the multiplication region up to the density $L_{sat}$
can be achieved, and the well-developed turbulence can be
realized. However, a large number of vortices cannot be created
for higher temperatures. In $^3$He-B, $q$ varies from zero, at
$T\rightarrow 0$, to $\infty$, at $T\rightarrow T_{\rm c}$, with
$q\sim 1$ at $T\sim 0.6T_{\rm c}$ \cite{Bevan}. As pointed out in
Ref.\ \cite{Nature03}, this temperature is in a good agreement
with the experimental observations of the onset of turbulence. In
superfluid He-II, the friction coefficient is always small except
for the immediate vicinity of $T_\lambda$. Thus, a an entangled
turbulent vortex state is usually formed. On the contrary, in
$^3$He-A the friction parameter $\alpha$ is large except for low
temperatures \cite{KL2tau} so that the turbulent state would be
possible only for very low temperatures.

To summarize, we develop a model for the onset of superfluid
turbulence taking into account the mutual friction between the
normal and superfluid components that appears in the presence of
quantized vortices. The condition for turbulent instability is
velocity independent: If the viscous component of the mutual
friction force is small, which happens at low temperatures, the
avalanche-like multiplication of seed vortex loops takes place.
The created vortex loops form a dense vortex tangle which further
evolves into a developed turbulent flow. On the contrary, the
number of created vortex loops is low and a turbulent state is not
formed in the viscosity dominated regime realized at higher
temperatures.

I am grateful to V. Eltsov, M. Krusius, and G. Volovik for many
stimulating discussions. This work was supported by Russian
Foundation for Basic Research.

\end{document}